\documentclass[english,twocolumn,showpacs,preprintnumbers,amsmath,amssymb,prb]{revtex4}
\usepackage[T1]{fontenc}
\usepackage[latin9]{inputenc}
\usepackage{amssymb}
\usepackage{graphicx}

\makeatletter
\@ifundefined{textcolor}{}
{%
 \definecolor{BLACK}{gray}{0}
 \definecolor{WHITE}{gray}{1}
 \definecolor{RED}{rgb}{1,0,0}
 \definecolor{GREEN}{rgb}{0,1,0}
 \definecolor{BLUE}{rgb}{0,0,1}
 \definecolor{CYAN}{cmyk}{1,0,0,0}
 \definecolor{MAGENTA}{cmyk}{0,1,0,0}
 \definecolor{YELLOW}{cmyk}{0,0,1,0}
}

\usepackage{epsfig}

\makeatother

\usepackage{babel}
\begin{document}

\title{Elastic properties and mechanical tension of graphene}

\author{R. Ram\'{\i}rez and C. P. Herrero}

\affiliation{Instituto de Ciencia de Materiales de Madrid (ICMM), Consejo Superior
de Investigaciones Cient\'{\i}ficas (CSIC), Campus de Cantoblanco,
28049 Madrid, Spain }
\begin{abstract}
Room temperature simulations of graphene have been performed as a
function of the mechanical tension of the layer. Finite-size effects
are accurately reproduced by an acoustic dispersion law for the out-of-plane
vibrations that, in the long-wave limit, behaves as $\rho\omega^{2}=\sigma k^{2}+\kappa k^{4}$.
The fluctuation tension $\sigma$ is finite ($\sim0.1$ N/m) even
when the external mechanical tension vanishes. Transverse vibrations
imply a duplicity in the definition of the elastic constants of the
layer, as observables related to the real area of the surface may
differ from those related to the in-plane projected area. This duplicity
explains the variability of experimental data on the Young modulus
of graphene based on electron spectroscopy, interferometric profilometery,
and indentation experiments. 
\end{abstract}


\maketitle

\section{Introduction}

Graphene is a solid surface in three-dimensional (3D) space.\citep{amorim16}
The area per atom, $A$, is a thermodynamic property difficult to
be measured. In fact the accessible observable is its projection,
$A_{p}$, onto the mean plane of the membrane, with $A_{p}\leq A$.
The equality is achieved in a strictly plane layer. The existence
of two different areas, $A$ and $A_{p}$, suggests a duplicity of
physical properties. For example, the negative thermal expansion coefficient
of graphene refers only to $A_{p}$, while the thermal expansion of
$A$ is positive.\citep{pozzo11,herrero16} An internal tension conjugated
to the actual membrane area $A$ should be distinguished from a mechanical
frame tension, $\tau$, conjugated to the projected area, $A_{p}$.
It is the tension $\tau$, the lateral force per unit length at the
boundary of $A_{p}$, the magnitude that defines the thermodynamic
ensemble in computer simulations.\citep{fournier08} $\tau$ is measurable
in fluid membranes by micropipette aspiration experiments.\citep{evans_90}
In addition, graphene elastic moduli, as the bulk or Young modulus,
may have different values if they are defined from fluctuations of
either $A$ or $A_{p}$. To avoid misunderstandings one should specify
unambiguously the kind of variable to which one is referring.

Differences between $A$ and $A_{p}$ originate from the existence
of ripples or wrinkles, that are a manifestation of the perpendicular
acoustic (ZA) vibrational modes of the layer. The \textit{harmonic}
long-wave limit ($k\rightarrow0$) of the ZA phonon dispersion is
$\rho\omega^{2}=\sigma k^{2}+\kappa k^{4}.$ Here $\rho$ is the atomic
mass density and $\kappa$ the bending rigidity of the layer. $\sigma$
is the fluctuation tension,\citep{fournier08,tarazona_13} that depends
on the applied mechanical tension as $\sigma=-\tau.$ \citep{pedro12a}
The anharmonicity of the out-of-plane fluctuations causes a renormalization
of the harmonic parameters. Room temperature simulations of free standing
graphene at\textit{ zero mechanical tension} $(\tau=0)$ reveal a
\textit{finite fluctuation tension} of $\sigma_{0}\sim0.1$ N/m.\citep{ramirez16}
This result agrees with analytical treatments of anharmonic effects
by perturbation theory,\citep{amorim14,michel15,adamyan16} with a
study of the coupling between vibrational and electronic degrees of
freedom by density functional calculations, \citep{kumar10} and with
the analysis of symmetry constraints in the phonon dispersion curves
of graphene.\citep{falkovsky08} All these studies are compatible
with an anharmonic relation between fluctuation and mechanical tensions
as $\sigma=\sigma_{0}-\tau.$ However, the long-wave limit predicted
by a membrane model with anomalous exponents deviates from this relation.\citep{los16a} 

In this paper, the anharmonicity of a free standing graphene layer
is studied by molecular dynamics (MD) simulations in the $N\tau T$
ensemble ($N$ being the number of atoms in the simulation cell and
$T$ the temperature). The fluctuation tension, $\sigma$, and the
bending rigidity, $\kappa$, of the layer are studied at $300$ K
as a function of both tensile $(\tau<0)$ and compressive $(\tau>0)$
stresses. The analytic long-wave limit, $\rho\omega^{2}=\sigma k^{2}+\kappa k^{4}$,
of the ZA phonons allows us the formulation of a finite-size correction
to the simulations. The amplitude of transverse fluctuations, $h^{2}$,
the projected area, $A_{p},$ and the bulk moduli, $B$ and $B_{p}$,
associated to the fluctuation of the areas $A$ and $A_{p}$, are
studied in the thermodynamic limit ($N\rightarrow\infty)$ as a function
of $\tau$. The bulk moduli ($B$ and $B_{p}$) are observables with
different behavior. While $B$ remains finite for all studied tensions,
$B_{p}\rightarrow0$ for a critical compressive tension, $\tau_{c}$.
This is the maximum tension that a planar layer can sustain, before
making a transition to a non-planar wrinkled structure. Our findings
provide light into the variability of experimental data on the Young
modulus of graphene based either on high-resolution electron energy
loss spectroscopy (HREELS) \citep{politano15}, on interferometric
profilometery,\citep{nicholl15} or on indentation experiments with
an atomic force microscope (AFM).\citep{lee08,lee13,lopez-polin15}

\section{Computational method}

\subsection{MD simulations}

The simulations are performed in the classical limit with a realistic
interatomic potential LCBOPII.\citep{los05} The original parameterization
was modified to increase the $T\rightarrow0$ limit of the bending
rigidity from $1.1$ eV to a more realistic value, $\kappa=1.5$ eV.\citep{ramirez16,lambin14}
A supercell $(N_{x},N_{y})$ of a 2D rectangular cell $(\mathbf{a},\mathbf{b})$
including 4 carbon atoms was employed with 2D periodic boundary conditions.\citep{ramirez16}
The supercell is chosen so that $N_{x}a\sim N_{y}b$. Runs consisted
of $10^{6}$ MD steps (MDS) for equilibration, followed by $8\times10^{6}$
MDS for the calculation of equilibrium properties. The time step amounts
to 1 fs. Full cell fluctuations were allowed in the $N\tau T$ ensemble.
Atomic forces were derived analytically by the derivatives of the
potential energy $U$. The stress tensor estimator was similar to
that used in previous works\citep{ramirez08c,ramirez16}
\begin{equation}
\tau_{xy}=\left\langle \frac{1}{A_{p}}\left(\sum_{i=1}^{N}mv_{ix}v_{iy}-\frac{\partial U}{\partial\epsilon_{xy}}\right)\right\rangle \;,
\end{equation}
where $m$ is the atomic mass, $v_{ix}$ is a velocity coordinate,
and $\epsilon_{xy}$ is a component of the 2D strain tensor. The brackets
$\left\langle \cdots\right\rangle $ indicates an ensemble average.
The derivative of $U$ with respect the strain tensor was performed
analytically. The mechanical tension is given by the trace of the
tensor
\begin{equation}
\tau=\frac{1}{2}\left(\tau_{xx}+\tau_{yy}\right)\:.
\end{equation}

The analyzed trajectories are subsets of $8\times10{{}^3}$ configurations
stored at equidistant times during the simulation run. The Fourier
analysis of transverse fluctuations was applied to simulations with
$N=960$ atoms to obtain $\sigma$ and $\kappa$ as a function of
$\tau$. Some simulations with $N=8400$ were performed to check the
convergence of the $\sigma$ and $\kappa$ calculation. The finite-size
effect in transverse fluctuations and projected area was studied with
additional simulations up to $N=33600$ atoms.

\subsection{Fourier analysis of the ZA modes}

The discrete Fourier transform of the heights of the atoms is
\begin{equation}
H_{ln}=\frac{1}{N}\sum_{j=1}^{N}h_{j}e^{-i\mathbf{k}_{ln}\mathbf{u}_{j}}\;.
\end{equation}
The position of the $j$'th atom is $\mathbf{r}_{j}=(\mathbf{u}_{j},h_{j})$,
where $\mathbf{u}_{j}$ is a 2D vector in the $(x,y)$ plane and the
height of the atom is $h_{j}$. Without loss of generality, the average
height of the layer is set as $\bar{h}=0.$ The wavevectors, $\mathbf{k}_{ln}$,
with wavelengths commensurate with the simulation supercell, are 
\begin{equation}
\mathbf{\mathbf{\mathbf{k}}}_{ln}=\left(\frac{l}{n_{k}}\frac{2\pi}{N_{x}a},\frac{n}{n_{k}}\frac{2\pi}{N_{y}b}\right)\:,\label{eq:k_n}
\end{equation}
with $l=0,\ldots,N_{x}-1$ and $n=0,\ldots,N_{y}-1$. $n_{k}$ is
an integer scaling factor to be defined below that unless otherwise
specified is identical to one. Assuming energy equipartition the mean-square
amplitude $\bar{H}_{ln}^{2}=H_{ln}H_{ln}^{*}$ of the ZA modes is
related to the phonon dispersion as
\begin{equation}
\left\langle \bar{H}_{ln}^{2}\right\rangle =\frac{k_{B}T}{A_{p}\rho\omega_{ln}^{2}}\;,\label{eq:A2}
\end{equation}
where $k_{B}$ is the Boltzmann constant. Our analysis of the long-wave
limit of $\left\langle \bar{H}_{ln}^{2}\right\rangle $ is reminiscent
of the simplest atomic model with an acoustic flexural mode, namely
a 1D chain of atoms with interactions up to second-nearest neighbors.
The dispersion relation for this model is\citep{ramirez16}
\begin{equation}
\rho\omega_{ln}^{2}=D\left[\sin^{2}\left(\frac{Lk_{ln}}{2}\right)-C\sin^{2}\left(Lk_{ln}\right)\right]\;.\label{eq:rho_w2}
\end{equation}
$k_{ln}$ is the module of the vector $\mathbf{\mathbf{\mathbf{k}}}_{ln}$.
The parameters ($D$, $L$, and $C$) are obtained by a least-squares
fit of the simulation results for $k_{ln}^{2}\left\langle \bar{H}_{ln}^{2}\right\rangle $
with the expression obtained by inserting Eq. (\ref{eq:rho_w2}) into
the r.h.s. of Eq. (\ref{eq:A2}) followed by multiplication by $k_{ln}^{2}$.
The fit is done for all wavevectors with $k_{ln}<10$ nm$^{-1}$.
The first two coefficients in the Taylor expansion of $\rho\omega^{2}$
as a function of $k$ provide $\sigma$ and $\kappa$ as\citep{ramirez16}
\begin{equation}
\sigma=DL^{2}\left(\frac{1}{4}-C\right)\;,
\end{equation}

\begin{equation}
\kappa=DL^{4}\left(\frac{C}{3}-\frac{1}{48}\right)\;.
\end{equation}

\section{Results and discussion}

The results of our MD simulations are divided into three Subsections
dealing with the long-wave limit of the ZA vibrations, the finite-size
correction of observables depending on the ZA modes, and the elastic
moduli of graphene.

\subsection{Long-wave limit of ZA modes}

\begin{figure}
\vspace{-0.5cm}
~\hspace{-1.0cm}
\includegraphics[width= 9.5cm]{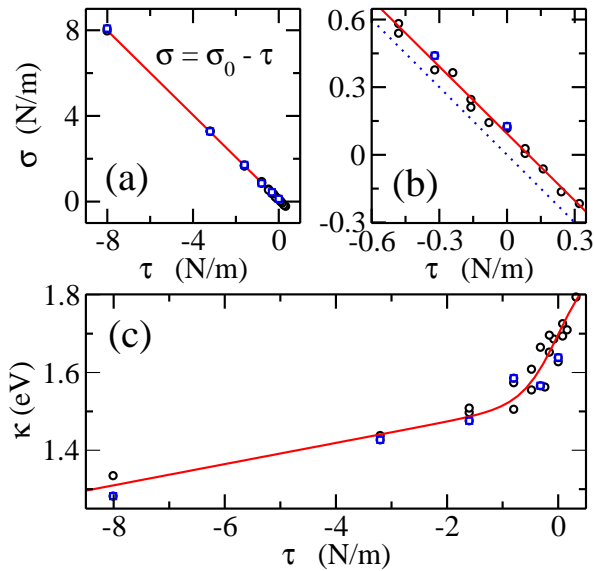}
\vspace{-0.8cm}
\caption{
(a) Dependence of the fluctuation tension, $\sigma,$ of graphene
with the mechanical tension $\tau$. Symbols are derived from $N\tau T$
simulations at 300 K with $N=960$ (black circles) and $N=8400$ (blue
squares). The full line is a linear fit; (b) Zoom of $\sigma$ for
small mechanical tensions. The dotted line is the harmonic expectation
$\sigma=-\tau$; (c) Results for the bending constant $\kappa$. The
full line shows a least-squares fit of the simulation data. }
\label{fig:sigma_kappa}
\end{figure}

The dependence of $\sigma$ and $\kappa$ with the mechanical tension
$\tau$ is displayed in Fig. \ref{fig:sigma_kappa}. $\tau$ varies
between 0.3 N/m, a value close to the maximum compressive stress ($\sim$0.5
N/m) sustained by a planar layer with $N=960$, and a tensile stress
of $-8$ N/m. The fluctuation tension obeys an anharmonic relation,
$\sigma=\sigma_{0}-\tau$, with $\sigma_{0}=0.094$ N/m. The value
of $\sigma$ in the vicinity of $\tau=0$ (see Fig. \ref{fig:sigma_kappa}b,
solid line) shows a clear shift from the harmonic expectation ($\sigma=-\tau$,
dotted line). $\kappa$ decreases monotonically as the mechanical
tension becomes more tensile (see Fig. \ref{fig:sigma_kappa}c). The
rate of decrease is smaller for tensions $\tau<-1$ N/m.

\subsection{Finite-size effects}

\begin{figure}
\vspace{-0.8cm}
~\hspace{-0.5cm}
\includegraphics[width= 9.5cm]{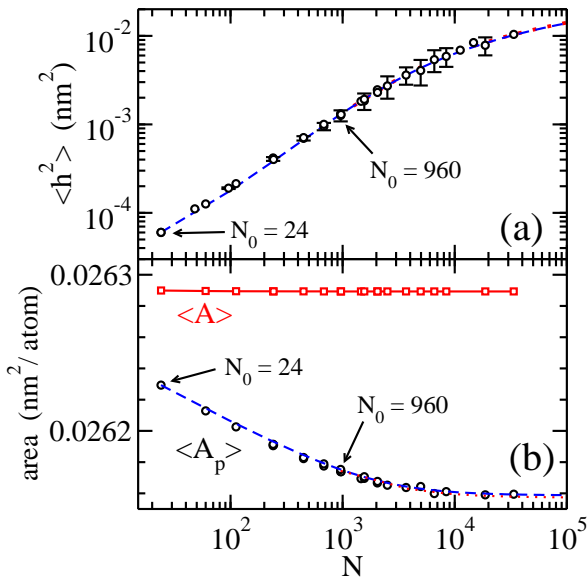}
\vspace{-0.8cm}
\caption{
$(a)$ Amplitude of the ZA modes as a function of the number of atoms
in the simulation cell. Symbols are $N\tau T$ simulation results
for $\left\langle h^{2}\right\rangle $ at 300 K and $\tau=0$. Broken
and dotted lines are finite-size corrections for $N_{0}=24$ and $960$
atoms, respectively. The two lines are nearly indistinguishable. $(b)$
Circles are the simulation results for the projected area $\left\langle A_{p}\right\rangle $.
The broken and dotted lines are finite-size extrapolations of the
simulations with $N_{0}=24$ and $960$ atoms, respectively. Squares
display the real area $\left\langle A\right\rangle $ from simulations
with $N=24$. The continuous line is a guide to the eye.}
\label{fig:h2_Ap}
\end{figure}

Finite-size effects are significant in graphene simulations.\citep{los16a}
The amplitude of the out-of-plane fluctuations,
\begin{equation}
\left\langle h^{2}\right\rangle =\frac{1}{N}\sum_{j=1}^{N}\left\langle h_{j}^{2}\right\rangle \:,
\end{equation}
is a function of $\sigma$ and $\kappa$, as these variables define
the long-wave limit of the ZA modes. Let us study the finite-size
error of the average $\left\langle h^{2}\right\rangle _{N_{0}}$ obtained
in a $N_{0}\tau T$ simulation. The $\mathbf{k}_{ln}$-grid in Eq.
(\ref{eq:k_n}) for the size $N_{0}$ is made up of elementary rectangles
$R_{i}$. Let $R_{0}$ be the rectangle having the $\Gamma$ point
at one vertex. The values of $(l,n)$ for the vertices of $R_{0}$
are (0,0), (0,1), (1,0), and (1,1), with (0,0) as the $\Gamma$-point.
Let us now consider successively larger cells defined with $N=N_{0}n_{k}^{2}$,
where $n_{k}=$$1,2,\ldots$ is an integer\textit{ scaling factor}.
Geometry in $\mathbf{k}$-space dictates that the larger the cell
size, the denser the $\mathbf{k}$-grid. The number of $\mathbf{k}$-points
in the elementary area $R_{0}$ increases as $(n_{k}+1)^{2}$, i.e.,
it grows as $2^{2},3^{2},4^{2},\ldots$ for $N$ increasing as $N_{0},2^{2}N_{0},3^{2}N_{0},\ldots$.
The finite-size correction for $\left\langle h^{2}\right\rangle _{N_{0}}$
is based on a discrete sum in reciprocal space. The sum is over the
$(n_{k}+1)^{2}$ $\mathbf{k}$-points in $R_{0}$ 
\begin{equation}
C_{N}=\frac{4}{N}\sum_{l=0}^{n_{k}}\sum_{n=0}^{n_{k}}{}^{'}\alpha_{ln}\bar{H}_{ln}^{2}\:.
\end{equation}
The prime indicates that the $\Gamma$-point ($l=n=0$) is excluded
from the sum. The multiplicative factor is the number of elementary
areas, $R_{0}$, in the Brillouin zone. It is equal to the multiplicity
of a general position in $k$-space, i.e., 4 (6) for a 2D rectangular
(hexagonal) unit cell. The amplitudes $\bar{H}_{ln}^{2}$ are calculated
by Eq. ($\ref{eq:A2}$) with the \textit{analytic} long-wave approximation
$\rho\omega_{ln}^{2}=\sigma k_{ln}^{2}+\kappa k_{ln}^{4}$. The weight
factors $\alpha_{ln}$ are unity except for those $\mathbf{k}_{ln}$
points at the vertices ($\alpha_{ln}=1/4)$ and sides ($\alpha_{ln}=1/2)$
of $R_{0}$. The finite-size correction to the average $\left\langle h^{2}\right\rangle _{N_{0}}$
is then
\begin{equation}
\left\langle h^{2}\right\rangle _{N}\approx\left\langle h^{2}\right\rangle _{N_{0}}+C_{N}-C_{N_{0}}\:.\label{eq:h2_correction}
\end{equation}
To check the reliability of this analytical model, we have compared
results for $\left\langle h^{2}\right\rangle _{N}$ derived from $N_{0}=24$
using Eq. (\ref{eq:h2_correction}), with those obtained directly
from simulations with $N$ atoms. Results for $\left\langle h^{2}\right\rangle $
at 300 K and $\tau=0$ with $N$ varying between 24 and 33600 atoms
are displayed in Fig. \ref{fig:h2_Ap}a as open circles. The finite-size
correction for $N_{0}=24$ is shown as a broken line. The agreement
with the simulation data is very good. Note that the average $\left\langle h^{2}\right\rangle =6\times10^{-5}$
nm$^{2}$ for $N=24$ increases by two orders of magnitude for $N=33600$.
The dispersion law $\rho\omega^{2}=\sigma k^{2}+\kappa k^{4}$ correctly
predicts the finite-size effect in $h^{2}$. The values of $\sigma$
and $\kappa$ at $\tau=0$ are 0.094 N/m and 1.7 eV (see Fig. \ref{fig:sigma_kappa}).
The finite-size correction obtained with $N_{0}=960$ is nearly indistinguishable
from that with $N_{0}=24$, an indication of the consistency of our
approach. The dispersion law, $\omega(k)$, implies that $\left\langle h^{2}\right\rangle $
increases with the size of the sample as $\ln N$.\citep{ramirez16}

A similar scheme applies to the size correction of $\left\langle A_{p}\right\rangle _{N_{0}}$.
Differential elements of the real and projected areas are related
by the surface metric as\citep{safran} 
\begin{equation}
dA=(1+h_{x}^{2}+h_{y}^{2})^{1/2}dA_{p}\approx\left(1+\frac{h_{x}^{2}+h_{y}^{2}}{2}\right)dA_{p}\:,\label{eq:dA}
\end{equation}
where $h_{x}$ ($h_{y}$) denotes the partial derivative of the height
$h$ with respect to $x$ ($y$), and the r.h.s. is a first-order
approximation when deviations from planarity are small. By integration
of the r.h.s and after Fourier transform one derives\citep{safran}
\begin{equation}
S_{N}=\frac{4}{N}\sum_{l=0}^{n_{k}}\sum_{n=0}^{n_{k}}{}^{'}\alpha_{ln}\frac{1}{2}k_{ln}^{2}\bar{H}_{ln}^{2}.\label{eq:C_Ap_N}
\end{equation}
The sum $S_{N}$ in reciprocal space provides the finite-size correction
to the projected area $\left\langle A_{p}\right\rangle _{N_{0}}$
as
\begin{equation}
\left\langle A_{p}\right\rangle _{N}\approx\left\langle A_{p}\right\rangle _{N_{0}}+S_{N}-S_{N_{0}}\:.\label{eq:Ap_N}
\end{equation}
Simulation results for the projected area $\left\langle A_{p}\right\rangle $
are presented in Fig. \ref{fig:h2_Ap}b as open circles. Size effects
are significant. $\left\langle A_{p}\right\rangle $ decreases with
increasing $N$, and converges to a finite area per atom for $N\rightarrow\infty$.
This behavior is in good agreement to previous Monte Carlo (MC) simulations
with the LCBOPII model.\citep{los16a} The finite-size correction
for $N_{0}=24$ using Eq. (\ref{eq:Ap_N}) is shown as a broken line.
A remarkable agreement to the simulation data is found. The correction
with $N_{0}=960$ is nearly indistinguishable from that with $N_{0}=24.$

Simulation results of the real area $\left\langle A\right\rangle $
with $N=24$ are shown in Fig. \ref{fig:h2_Ap}b. $A$ is calculated
by triangulation of the surface, with C atoms and hexagon centers
as vertices of the triangles. Hexagon centers are located at the average
position of their six vertices. The area $\left\langle A\right\rangle $,
in contrast to $\left\langle A_{p}\right\rangle $, displays a small
finite-size error, not visible at the scale of Fig. \ref{fig:h2_Ap}b.
For $N=24$, the relative finite-size error in $\left\langle A\right\rangle $
amounts to $2\times10^{-3}$ \%, while that of $\left\langle A_{p}\right\rangle $
is two orders of magnitude larger, $0.3$ \%. A larger size, $N\sim4\times10^{4}$,
is needed to reduce the finite-size error of $\left\langle A_{p}\right\rangle $
to the small error achieved for $\left\langle A\right\rangle $ with
$N=24$. Note that our finite-size correction considers only the acoustic
ZA mode. The obtained results imply that the rest of vibrational modes
(namely the in-plane and optical out-of-plane (ZO) modes of the layer)
display a comparatively small size effect. 

The difference between $A$ and $A_{p}$ for a \textit{continuous}
membrane in the $N\rightarrow\infty$ limit can be calculated by integration
and Fourier transform of the r.h.s. of Eq. (\ref{eq:dA}).\citep{safran,chacon15}
With the ZA dispersion law, $\rho\omega^{2}=\sigma k^{2}+\kappa k^{4}$,
one gets
\begin{eqnarray}
A = A_{p} \left(1+\frac{k_{B}T}{4\pi}\int_{0}^{k_{max}}dk\frac{k}
         {\sigma+\kappa k^{2}}\right) =   \nonumber \\
    = A_{p}\left[1+\frac{k_{B}T}{8\pi\kappa}\ln\left(1+\frac{2\pi\kappa}
        {\sigma A_{p}}\right)\right]\:,
\end{eqnarray}
with $k_{max}=\left(2\pi/A_{p}\right)^{1/2}$. A quadratic term $\sigma k^{2}$
in $\rho\omega^{2}$ is a sufficient condition for the convergence
of the integral.

\subsection{Elastic moduli}

\begin{figure}
\vspace{-0.5cm}
~\hspace{-0.5cm}
\includegraphics[width= 9.5cm]{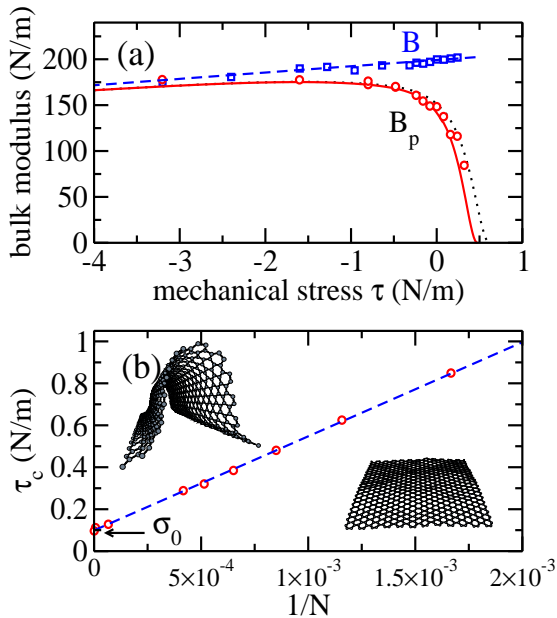}
\vspace{-0.8cm}
\caption{
$(a)$ Bulk moduli of graphene at 300 K as a function of the mechanical
tension $\tau$. $B$ and $B_{p}$ are defined with respect to the
real and projected areas, $A$ and $A_{p},$ respectively. Circles
are simulation results of $B_{p}$ for $N=960$. The dotted and continuous
lines are extrapolations of $B_{p}$ for $N=864$ and $N=1176$ atoms,
respectively. Both lines were derived by finite-size correction of
simulations with $N_{0}=24$ atoms. Squares are simulation results
of $B$ for $N=24$ atoms. The broken line is a linear fit. $(b)$
Critical mechanical tension as a function of the number of atoms.
Results derived by finite-size correction of simulations with $N_{0}=24$
atoms. Typical atomic structures with $N=960$ are shown below (planar
layer) and above (wrinkled layer) the critical tension.}
\label{fig:B}
\end{figure}

We focus now on the elastic moduli of graphene. First, the finite-size
correction $\left\langle A_{p}\right\rangle _{N}$ is derived with
$N_{0}=24$ at mechanical tensions $\tau$ in the range $-4$ to $0.05$
N/m, using the values of $\sigma$ and $\kappa$ shown in Fig. \ref{fig:sigma_kappa}.
For each tension $\tau$ and size $N$, $\left\langle A_{p}\right\rangle _{N}$
is then obtained at two close tensions $\tau\pm$0.016 N/m, in order
to calculate numerically the derivative $\partial\left\langle A_{p}\right\rangle _{N}/\partial\tau$.
The bulk modulus $B_{p}$ for size $N$ is then obtained as 
\begin{equation}
B_{p}=-\left\langle A_{p}\right\rangle _{N}\frac{\partial\tau}{\partial\left\langle A_{p}\right\rangle _{N}}\:.\label{eq:B_der}
\end{equation}
The values of $B_{p}$ for the sizes $24n_{k}^{2}$, with $n_{k}=6$
($N=864$) and $n_{k}=7$ ($N=1176$ ) are plotted in Fig. \ref{fig:B}a
as dotted and full lines, respectively. For comparison, open circles
display $B_{p}$ from $N\tau T$ simulation with $N=960$ atoms, as
derived from the fluctuation formula\citep{herrero08}
\begin{equation}
B_{p}=\frac{k_{B}T\left\langle A_{p}\right\rangle }{N\left(\left\langle A_{p}^{2}\right\rangle -\left\langle A_{p}\right\rangle ^{2}\right)}\:.
\end{equation}
It is remarkable the agreement between the values of $B_{p}$ from
the simulations with $N=960$ atoms and from the finite-size extrapolation
with $N_{0}=24$. This agreement is more demanding than that of $\left\langle h^{2}\right\rangle $
and $\left\langle A_{p}\right\rangle $ in Fig. \ref{fig:h2_Ap},
because of the wide range of studied mechanical tensions. 

The bulk modulus, $B$, calculated from the fluctuation of the real
area, $\left\langle A\right\rangle $, in simulations with $N=24$
is shown as open squares in Fig. \ref{fig:B}. The finite-size effect
in $B$ is negligible at the scale of Fig. \ref{fig:B}a, in line
with the negligible finite-size effect in the real area $\left\langle A\right\rangle $
(see Fib. \ref{fig:h2_Ap}b). $B$ and $B_{p}$ behave quite differently.
The anharmonicity causes a finite derivative of $B$ with respect
to $\tau$, $B'\sim7.$ $B_{p}$ is close to $B$ at the largest studied
negative tensions, when out-of-plane fluctuations are small, but $B_{p}$
becomes much smaller than $B$ as $\tau$ increases. At a critical
compressive tension, $\tau_{c}>0,$ the bulk modulus $B_{p}$ vanishes.
$\tau_{c}$ represents the stability limit for a planar layer before
the stable configuration becomes wrinkled. $\tau_{c}$ displays a
strong size effect that is shown in Fig. \ref{fig:B}b. The critical
mechanical tension, $\tau_{c},$ decreases with the number of atoms
as $N^{-1}$. In the thermodynamic limit we get $\tau_{c}=\sigma_{0}$,
i.e., a planar layer is able to sustain a compressive tension of about
$\tau_{c}\sim0.1$ N/m before becoming wrinkled. The structural plots
in Fig. \ref{fig:B}b shows that wrinkles are generated along a preferential
direction. 

\begin{figure}
\vspace{-1.0cm}
\includegraphics[width= 8.5cm]{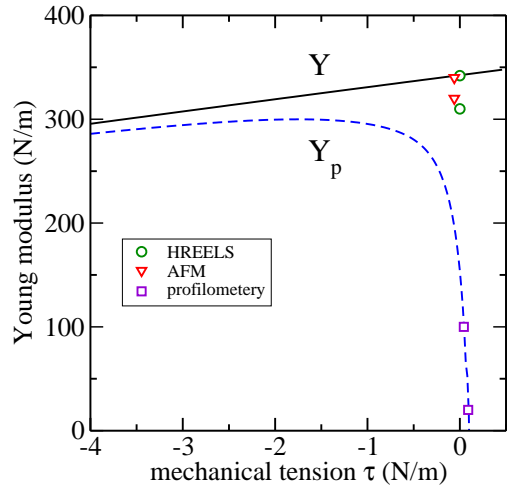}
\vspace{-0.8cm}
\caption{
The Young moduli, $Y$ and $Y_{p},$ are displayed at 300 K as a function
of the mechanical tension $\tau$. $Y$ and $Y_{p}$ are defined with
respect to the real and projected areas, $A$ and $A_{p},$ respectively.
$Y_{p}$ was derived by finite-size correction of the simulation with
$N_{0}=24$ atoms. $Y$ displays a very small size effect and it was
derived by a least-squares fit of simulations with $N=24$ atoms.
The Poisson ratio is $\nu=0.15$. Symbols are experimental values
of the Young modulus of graphene as measured by HREELS,\citep{politano15}
AFM, \citep{lee08,lee13,lopez-polin15} and interferometric profilometery.\citep{nicholl15} }
\label{fig:Y}
\end{figure}

The Young modulus, $Y$, of a 2D layer is related to the bulk modulus
by $Y=2B(1-\nu)$, where $\nu$ is the Poisson ratio. We have calculated
$\nu=0.15$ for the employed LCBOPII model in the classical $T\rightarrow0$
limit. Using this value, the Young moduli, $Y$ and $Y_{p}$, of graphene
have been plotted in Fig. \ref{fig:Y} as a function of $\tau$. $Y_{p}$
was derived in the thermodynamic limit ($N\rightarrow\infty$) by
applying the finite-size correction to simulations with $N_{0}=24$.
$Y$ has a small size effect and it was derived by a least-squares
fit of simulations with $N=24$ atoms. The Young modulus $Y$, related
to the real area $A$, shows a monotonic dependence with $\tau$.
For $\tau=0$ we find $Y=339$ N/m. On the other side, $Y_{p}$ displays
a maximum ($297$ N/m) at $\tau=-1.7$ N/m, decreases rapidly for
$\tau\gtrsim-1$ N/m and vanishes at the critical tension $\tau_{c}=0.1$
N/m. 

Experimental HREELS results of the Young modulus of both planar and
corrugated graphene supported on a variety of metal substrates are
displayed in Fig. \ref{fig:Y} as open circles.\citep{politano15}
HREELS provides in-plane phonon dispersion curves. The elastic constants
derived from the sound velocities of the acoustic in-plane branches
are a property related to the real area of the layer that should correspond
to the observable $Y$. In fact, we find good agreement between the
HREELS results and our simulation results for $Y$. Results from AFM
indentation experiments, shown as triangles in Fig. \ref{fig:Y},
also agree with our simulation results for $Y$.\citep{lee08,lee13,lopez-polin15}
The lack of correlation between the elastic modulus and the mechanical
tension, reported in the experiments of Ref. \onlinecite{lopez-polin15},
is in line with the weak dependence of the simulation results of $Y$
on the value of $\tau$. However, elastic constants from interferometric
profilometery are derived by fitting the experimental data to an \textit{average
surface}.\citep{nicholl15} These elastic constants, plotted as squares
in Fig. \ref{fig:Y}, should correspond to the observable $Y_{p}$.
The two interferometric profilometery results are displayed at mechanical
tensions of 0.04 and 0.09 N/m. The tension of the graphene layer depends
on the sample processing by factors that can not be controlled experimentally.
Thus the previous tensions have been chosen to fit the experimental
data to our $Y_{p}$ curve.

\section{Summary}

We have analyzed the long-wave limit of the acoustic transverse fluctuations
of graphene at 300 K. A finite-size correction for the out-out-plane
amplitude, $h^{2}$, the projected area, $A_{p}$, and the bulk modulus,
$B_{p}$, has been based on the dispersion relation, $\rho\omega^{2}=\sigma k^{2}+\kappa k^{4}$.
The size correction has small computational cost, displays excellent
agreement to simulations with larger cells, and strongly supports
the validity of the acoustic dispersion law in graphene. The fluctuation
tension, $\sigma$, depends on the external mechanical tension, $\tau$,
by an anharmonic relation, $\sigma=\sigma_{0}-\tau$. At 300 K we
find $\sigma_{0}\sim0.1$ N/m. The finite value of $\sigma_{0}$ has
a large influence in the amplitude of the out-of-plane fluctuations
and in the mechanical stability of the crystalline membrane against
wrinkling. The Young modulus, $Y_{p}$, related to the projected area
varies between 0 and $\sim300$ N/m depending upon the mechanical
tension sustained by the layer. However, the Young modulus, $Y$,
related to the real area, amounts to $340$ N/m in the absence of
external mechanical tension, and decreases to $\sim300$ N/m for large
tensile stresses of -4 N/m. The existence of two different observables,
$Y$ and $Y_{p}$, provides a reliable explanation for the experimental
values of the Young modulus of graphene as measured by HREELS, AFM,
and interferometric profilometery.

\acknowledgments

This work was supported by Direcci\'on General de Investigaci\'on, MINECO
(Spain) through Grants No. FIS2012-31713, and FIS2015-64222- C2-1-P.
We thank the support of J. H. Los in the implementation of the LCBOPII
potential.

\end{document}